\input{psfig}
\documentstyle[onecolumn,doublespacing]{mn}
\newif\ifAMStwofonts
\ifoldfss
  \ifCUPmtlplainloaded \else
    \NewTextAlphabet{textbfit} {cmbxti10} {}
    \NewTextAlphabet{textbfss} {cmssbx10} {}
    \NewMathAlphabet{mathbfit} {cmbxti10} {} % for math mode
    \NewMathAlphabet{mathbfss} {cmssbx10} {} %  "   "    "
  \fi
  \ifAMStwofonts
    \ifCUPmtlplainloaded \else
      \NewSymbolFont{upmath} {eurm10}
      \NewSymbolFont{AMSa} {msam10}
      \NewMathSymbol{\upi}     {0}{upmath}{19}
      \NewMathSymbol{\umu}     {0}{upmath}{16}
      \NewMathSymbol{\upartial}{0}{upmath}{40}
      \NewMathSymbol{\leqslant}{3}{AMSa}{36}
      \NewMathSymbol{\geqslant}{3}{AMSa}{3E}

       \let\ge=\geqslant
    \fi
  \fi
\fi % End of OFSS

\ifnfssone
  \newmathalphabet{\mathit}
  \addtoversion{normal}{\mathit}{cmr}{m}{it}
  \addtoversion{bold}{\mathit}{cmr}{bx}{it}
  \newmathalphabet{\mathbfit} % math mode version of \textbfit{..}
  \addtoversion{normal}{\mathbfit}{cmr}{bx}{it}
  \addtoversion{bold}{\mathbfit}{cmr}{bx}{it}
  \newmathalphabet{\mathbfss} % math mode version of \textbfss{..}
  \addtoversion{normal}{\mathbfss}{cmss}{bx}{n}
  \addtoversion{bold}{\mathbfss}{cmss}{bx}{n}
  \ifAMStwofonts
    \ifCUPmtlplainloaded \else
      \UseAMStwoboldmath
      \makeatletter
      \new@mathgroup\upmath@group
      \define@mathgroup\mv@normal\upmath@group{eur}{m}{n}
      \define@mathgroup\mv@bold\upmath@group{eur}{b}{n}
      \edef\UPM{\hexnumber\upmath@group}
      \new@mathgroup\amsa@group
      \define@mathgroup\mv@normal\amsa@group{msa}{m}{n}
      \define@mathgroup\mv@bold\amsa@group{msa}{m}{n}
      \edef\AMSa{\hexnumber\amsa@group}
      \makeatother
      \mathchardef\upi="0\UPM19
      \mathchardef\umu="0\UPM16
      \mathchardef\upartial="0\UPM40
      \mathchardef\leqslant="3\AMSa36
      \mathchardef\geqslant="3\AMSa3E

       \let\ge=\geqslant
    \fi
  \fi
\fi % End of NFSS release 1

\ifnfsstwo
  \DeclareMathAlphabet{\mathbfit}{OT1}{cmr}{bx}{it}
  \SetMathAlphabet\mathbfit{bold}{OT1}{cmr}{bx}{it}
  \DeclareMathAlphabet{\mathbfss}{OT1}{cmss}{bx}{n}
  \SetMathAlphabet\mathbfss{bold}{OT1}{cmss}{bx}{n}
  \ifAMStwofonts
    \ifCUPmtlplainloaded \else
      \DeclareSymbolFont{UPM}{U}{eur}{m}{n}
      \SetSymbolFont{UPM}{bold}{U}{eur}{b}{n}
      \DeclareSymbolFont{AMSa}{U}{msa}{m}{n}
      \DeclareMathSymbol{\upi}{0}{UPM}{"19}
      \DeclareMathSymbol{\umu}{0}{UPM}{"16}
      \DeclareMathSymbol{\upartial}{0}{UPM}{"40}
      \DeclareMathSymbol{\leqslant}{3}{AMSa}{"36}
      \DeclareMathSymbol{\geqslant}{3}{AMSa}{"3E}

       \let\ge=\geqslant
    \fi
  \fi
\fi % End of NFSS release 2

\ifCUPmtlplainloaded \else
  \ifAMStwofonts \else % If no AMS fonts
    \def\upi{\pi}
    \def\umu{\mu}
    \def\upartial{\partial}
  \fi
\fi

\title{Profiles of dark-matter haloes at high redshift}
\author[A. Del Popolo]
       {A. Del Popolo \\
Dipartimento di Matematica, Universit\`{a} Statale di Bergamo,
Piazza Rosate, 2 - I 24129 Bergamo, ITALY\\
Feza G\"ursey Institute, P.O. Box 6 \c Cengelk\"oy, Istanbul, Turkey\\
Bo$\breve{g}azi$\c{c}i University,  Physics Department,
80815 Bebek, Istanbul, Turkey
}
\date{}

\pagerange{\pageref{firstpage}--\pageref{lastpage}}
\pubyear{}

\begin{document}

\maketitle

\label{firstpage}

\begin{abstract}
I study the evolution of haloes density profiles as a function of time in
the SCDM and $\Lambda$CDM cosmologies.
%Supposing that the density profiles
%of halos at all time 
Following Del Popolo et al. (2000) (hereafter DP2000), I calculate the concentration
parameter $c=r_{v}/a$ and study its time evolution.
%study the evolution of the concentration parameter $c(z)=r_{200}(z)/a(z)$
%in the SCDM and $\Lambda$CDM cosmology by means
%of the improved SIM (secondary infall model)
%introduced in Del Popolo et al. (1999).
For a given halo mass, I find that 
$c(z) \propto 1/(1+z)$ both in the $\Lambda$CDM and
SCDM cosmology, in agreement with
the analytic model of Bullock et al. (1999) (hereafter B99)
and N-body simulations.
%%and
%%$c(z) \propto \simeq \Omega(z)^{1/3}/(1+z)$ in the $\Lambda$CDM model,
%%that for $z \ge 1$ is in agreement with 
%%the N-body simulations of Bullock et al. (1999).
In both models,
$a(z)$ is roughly constant.
The present model predicts a stronger evolution of $c(z)$ with respect
to Navarro, Frenk \& White (1997) (hereafter NFW97) model.
%
%
%%The comparison of $c(z)$ given by the present
%%model with that of
%%Navarro et al. (1997) (hereafter NFW97)
%%shows that this last model predicts a smaller
%%variation in time with respect to our.
%
%We also compared our $c(z)$ to that obtained by Bullock et al. (1999)
%%The prediction for $c(z)$ is in agreement
%%with the N-body simulations of Bullock et al. (1999) in the case of 
%%$\Lambda$CDM (they only simulated this cosmology) for $z \ge 1$
%%and reproduces exactly the result of their analytic model in the SCDM case.
Finally I show some consequences of the results on galaxy modelling.
\end{abstract}


\section{Introduction}

The structure of dark matter haloes is of fundamental importance in the
study of
the formation and evolution of galaxies and
clusters of galaxies. From the theoretical point of view, the structure of
dark matter haloes can be studied both analytically and numerically.
A great part of the
analytical work done so far is based on the secondary infall model (SIM)
introduced by Gunn \& Gott (1972).
Calculations based on this model predict that the density profile of
the virialized halo should scale as $\rho \propto r^{-9/4}$.
Self-similar solutions were found by Fillmore \& Goldreich (1984) and
Bertschinger (1985), who obtained a profile of $\rho\propto r^{-2.25}$.
Hoffman \& Shaham (1985) (hereafter HS) considered a scale-free initial
perturbation spectra, $P(k) \propto k^n$.
They showed that
$\rho\propto {r^{-\alpha}}$ with
$\alpha=\frac{3(3+n)}{(4+n)}$, thus recovering Bertschinger's (1985)
profile for $n=0$ and
$\Omega=1$.
They also showed that, in an open Universe, the slopes of
the density profiles steepen with increasing values of $n$ and
with decreasing $\Omega$, reaching a profile
$\rho\propto r^{-4}$ for $\Omega \rightarrow 0$.\\
N-body simulations,
such as that of Quinn, Salmon \& Zurek (1986), 
West, Dekel \& Oemler (1987) and Efstathiou et al. (1988) arrived at conflicting
results implying that better
numerical resolution were needed to settle the issue. Recent results from
higher resolution simulations (Navarro, Frenk \& White 1995, 1996, 1997 (hereafter
NFW95, NFW96, NFW97); Lemson 1995; Cole \& Lacey 1996; Tormen, Diaferio \& Syer 1998)
obtained, using different codes and different setups for the
initial conditions, agreement in the conclusion
%%(for sake of precision we remember that
%%these last results
%%are not universally accepted:
%%see Klypin et al. (1997) and Nusser \& Shet (1998))
that halo density profiles
do not follow a power law but
develope a universal profile,
a one parameter functional form that provides a good fit to haloes over
a large range of masses and for 
%
%
%a quite general profile
any scenario in which structures form due to hierarchical
clustering, characterized by a slope $\beta=\frac{d \ln \rho}{d ln r}=-1$ near
the halo centre and $\beta=-3$ at large radii.
In that approach, density profiles can be fitted with the functional form:
\begin{equation}
\frac{\rho(r)_{\rm NFW}}{\rho_{\rm b}}= \frac{\delta_{\rm n}}{\frac{r}{a}\left(1+\frac{r}{a}\right)^2}
\label{eq:nfw}
\end{equation}
where $\rho_{\rm b}$ is the background density and $ \delta_{\rm n}$ is the
central overdensity [below I shall refer to equation (\ref{eq:nfw}) (NFW97)
as the NFW profile].
The scale radius $a$, which defines the scale where the
profile shape changes from slope $\beta<-2$ to $\beta>-2$,
and the characteristic overdensity, $\delta_{\rm n}$, are related
%by:
%\begin{equation}
%\delta_c=\frac{200}{3} \frac{c^3}{}
%\end{equation}
because the
mean overdensity enclosed within the virial radius
$r_{\rm v}$ is $ \simeq 180$.
I recall that according to NFW96, 97), $a$ is linked to a
"concentration" parameter, $c$, by the relation $a=\frac{r_{\rm v}}{c}$ and
the
parameter $c$ is linked to the characteristic density, $\delta_{\rm n}$,
by the relation:
\begin{equation}
\delta_{\rm n}=\frac{200}{3} \frac{c^3}{\ln(1+c)-c/(1+c)}
\label{eq:dn}
\end{equation}
The scale radius and the central overdensity are directly related to the
formation time of a given halo (NFW97). The power spectrum
and the cosmological parameters only enter to determine the typical formation
epoch of a halo of a given mass, and thereby the dependence of the
characteristic radius, $a$, or the overdensity
$\delta_{\rm n}$ on the total mass of the halo. $\delta_{\rm n}$ increases
for decreasing virial mass,
$M_{\rm v}$. A natural reason for the fact that low-density haloes tend
to show higher densities is that they typically collapse earlier, when
the universe was denser. To model this trend, NFW97 proposed a step-by-step
calculation of the density profile assuming that the characteristic density,
$\delta_{\rm n}$, is proportional to the density of the universe at the
corresponding collapse redshift, $z_{\rm c}$.
This model successfully predicts the $\delta_{\rm n}-M_{\rm v}$ relation
for different cosmological models at $z=0$. The model has been also extended
in NFW97 to predict the redshift dependence of the halo
profile parameters but, as shown by B99 for
the $\Lambda$CDM
cosmology, the evolution of $c$ (and consequently of $\delta_{\rm n}$) is much
stronger than in the NFW97 model. \\
In this paper, I use the improved SIM introduced by DP2000
to determine $c(z)$ for both SCDM and $\Lambda$CDM models
and to compare it with prediction of NFW97 and B99 models. \\
%%and with that of Bullock et al. (1999). \\
The plan of the paper is the following: in Section ~2, I introduce the 
%Del Popolo et al. (1999)
model. In Section ~3, I show the results of the model
and Section ~4 is devoted to the conclusions.

\section[]{Time evolution of the concentration parameter}

The simplest version of SIM considers an initial point mass, which acts
as a nonlinear seed, surrounded by a homogeneous uniformly expanding
universe. Matter around the seed slows down due to its gravitational
attraction, and eventually falls back in concentric spherical shells with
pure radial motions. The assumptions of SIM that are most often questioned
are the spherical symmetry and the absence of peculiar
velocities (non-radial motions): in the "real" collapse, accretion does not
happen in spherical shells but by aggregation of subclumps of matter which
have already collapsed; a large fraction of observed clusters of galaxies
exhibit significant substructure (Kriessler et al. 1995). Motions are not
purely radial, especially when the perturbation detaches from the general
expansion. Nevertheless the SIM gives good results in describing the formation
of dark matter haloes, because in energy space the collapse is ordered
and gentle, differently from what is seen in N-body simulations
(Zaroubi, Naim \& Hoffman 1996). As I showed in a recent paper, DP2000,
the discrepancies between the SIM and some high resolution N-body
simulations are not due to the spherical symmetry assumption
%the previous quoted intrinsic limits
of the SIM but arises because of some non-accurate assumptions
used in its implementation.
As I showed in DP2000, the predictive power of the
SIM is greatly improved when some problems of the previous implementations
are removed. \\
To begin with, the conclusion $\rho \propto r^{-2}$ for $n <-1$, claimed
by HS, 
is not a direct consequence of the HS
model, but it is an assumption made by the quoted authors,
following the study of
self-similar gravitational collapse by Fillmore \& Goldreich (1984). In fact,
as reported by the same authors,
in deriving the relation between the density at maximum expansion and
the final one, HS assumed that each new shell that collapses
can be considered as a small perturbation to the gravitational
field of the collapsed halo. This assumption breaks down for $n<-1$. \\
Secondly, the assumption made by Hoffman \& Shaham (1985) that
$\delta(r) \propto \xi(r) \propto r^{-(3+n)}$ is not good for regions internal
to the virial radius, $r_{\rm v}$
(see Peebles 1974; Peebles \& Groth 1976;
Davis \& Peebles 1977; Bonometto \& Lucchin 1978; Peebles 1980; Fry 1984).
In the inner regions of the halo, scaling arguments plus the stability
assumption tell us that $\xi(r) \propto r^{-\frac{3(3+n)}{(5+n)}}$, and
we expect a slope different from that of HS.
In other words, HS's (1985) solution applies only to
the outer regions of collapsed haloes, and consequently the conclusion,
obtained from that model,
that dark matter haloes density profiles
can be approximated by power-laws on their overall radius range
is not correct. It is then necessary to
introduce a model that can make predictions also on the inner parts of
haloes. \\
Thirdly, according to Bardeen et al. (1986),
(hereafter BBKS), the mean
peak profile depends on a sum involving the initial correlation function,
$\xi(r) \propto r^{-(3+n)}$,
and its Laplacian, ${\bf \bigtriangledown}^2 \xi(r) \propto r^{-(5+n)}$
(BBKS; Ryden \& Gunn 1987):
\begin{equation}
\delta (r)  =\frac{\nu \xi (r)}{\xi (0)^{1/2}}-\frac{\vartheta
(\nu ,\gamma )}{\gamma (1-\gamma ^2)}\left[ \gamma ^2\xi (r)+\frac{%
R_{\ast }^2}3\nabla ^2\xi(r) \right] \cdot \xi (0)^{-1/2}
\label{eq:dens}
\end{equation}
where $\nu $ is the height of a density peak:
\begin{equation}
\nu=\frac{\delta(0)}{\sigma(R,z)}
\end{equation}
The variance $\sigma(R,z)$ is given by:
\begin{equation}
\sigma ^2(R,z)=D^2(z,\Omega) \frac 1{2\pi ^2}\int_0^\infty dkk^2P(k)W^2(kR)
\label{eq:ma3}
\end{equation}
where the function $D(z,\Omega)$ describes the growth of
density fluctuations (Peebles 1980) and $W(kR)$ is a top-hat smoothing function:
\begin{equation}
W(kR)=\frac 3{\left( kR\right) ^3}\left( \sin kR-kR\cos kR\right)
\label{eq:ma4}
\end{equation}
%$\xi (r)$ is the two-point
%%correlation function
$\gamma $ and $R_{\ast}$ are two spectral parameters
given respectively by:
\begin{equation}
\gamma =\frac{\int k^4P(k)dk}{\left[ \int k^2P(k)dk\int k^6P(k)dk\right]
^{1/2}}
\end{equation}
\begin{equation}
R_{*}=\left[ \frac{3\int k^4P(k)dk}{\int k^6P(k)dk}\right] ^{1/2}
\end{equation}
while $ \vartheta (\gamma \nu ,\gamma )$ is:
\begin{equation}
\vartheta (\nu \gamma ,\gamma )=\frac{3(1-\gamma ^2)+\left( 1.216-0.9\gamma
^4\right) \exp \left[ -\left( \frac \gamma 2\right)
\left( \frac{\nu \gamma }%
2\right) ^2\right] }{\left[ 3\left( 1-\gamma ^2\right) +0.45+\left( \frac{%
\nu \gamma }2\right) ^2\right] ^{1/2}+\frac{\nu \gamma }2}
\label{eq:tet}
\end{equation}
I recall that the $z$ dependence of $\delta$ is:
\begin{equation}
\delta(z)=\delta_0 D(z,\Omega)
\end{equation}
being $\delta_0$ the overdensity as measured at current epoch $t_0$.
As can be seen for example in the case of a scale-free density perturbation
spectrum (DP2000, equation (20)), the initial mean density obtained
using the model of this paper
is extremely different from that obtained and used in HS. \\
The first step to get $c(z)$ is to calculate $\delta(r)$ for a given
cosmology starting from the related spectrum.
%%I calculate $\delta(r)$, and $c(z)$,
In order to calculate $\delta(r)$ in the SCDM cosmology
($\Omega=1$, $h=0.5$, $n=1$), I use the spectrum given by BBKS:
\begin{eqnarray}
P(k) = Ak^{-1}\left[ \ln \left( 1+4.164k\right) \right] ^2 \nonumber\\
\times \left(192.9+1340k+1.599\times 10^5k^2+1.78\times 10^5k^3+3.995\times
10^6k^4\right) ^{-1/2}
%e^{-k^2l^2/2}
\end{eqnarray}
normalized by
imposing that the mass variance at $8h^{-1}$Mpc is $\sigma _{8}=0.63$.
For 
the $\Lambda$CDM model ($\Omega_{\rm m}=0.3$, $\Omega_{\Lambda}=0.7$,
$h=0.7$), I also use the BBKS spectrum normalized as
$\sigma_8=1$. 
Supposing that energy is conserved,
the shape of the density profile at maximum of expansion is
conserved after the virialization,
and
%the relaxed density profile
is given by (Peebles 1980; HS; White \& Zaritsky 1992):
\begin{equation}
\rho(r)=\rho_{\rm i}
%%\frac{(1+z)^3}{\Omega(z)}
\left( \frac{r_{\rm i}}{r} \right)^2 \frac{d r_{\rm i}}{dr}
\end{equation}
where $r_{\rm i}$ and $\rho_{\rm i}$ are respectively the initial radius and
the density, while $\Omega(z)$ is the density parameter at epoch $z$.
The final radius, $r$, and the initial one, $r_{\rm i}$ are connected by:
\begin{equation}
r=F r_{\rm m} =F r_{\rm i}\frac{1+{\overline
\delta_{\rm i}}}{{\overline \delta_{\rm i}}-(\Omega_{\rm i}^{-1}-1)}
\label{eq:rv}
\end{equation}
where $r_{\rm m}$ is the shell radius at maximum expansion, $F$ is given in
DP2000 (equation (26)) and the mean
fractional density excess inside a given radius, $\overline{\delta}$, is: 
\begin{equation}
{\overline \delta}=\frac{3}{r^3} \int_0^r \delta(y)y^2 dy
\end{equation}
In order to calculate the evolution of $c=r_{\rm v}/a$, I must calculate
the inner radius $a$ and the virial radius, $r_{\rm v}$.
The inner radius, $a$, is characterized by the condition:
\begin{equation}
\frac{d log \rho(r)}{d log (r)}|_{a}=-2
\end{equation}
%The density parameter, $\delta_n$, is related to the NFW density at
%$a$ by $\delta_n=4 \rho_{\rm NFW}$
while the virial radius, $r_{\rm v}$, is
the radius within which the mean overdensity is $\delta_{\rm v}$
times the  critical density, $\rho_{\rm c}$, at that redshift:
%
%%mean
%%universal density, $\rho_{\rm b}$, at that redshift:
%
\begin{equation}
M_{\rm v}= \delta_{\rm v}(z) \rho_{\rm c}(z) \frac{4 \pi}{3} r_{\rm v}(z)^3
\label{eq:mass}
\end{equation}
where $M_{\rm v}$ is the virial mass of the halo
%%the virial overdensity,
%%$\delta_{\rm v}$,
%%is provided by the spherical top-hat collapse model
and the critical density
$\rho_{\rm c}$ is:
\begin{equation}
%%\rho_{\rm b} (z)=\rho_{0 {\rm b}}
%%\left[\Omega_0 (1+z)^3 +\Omega_{\rm R}(1+z)^2+(1-\Omega_0)
%%\right]
\rho_{\rm c} (z)=\rho_{0 {\rm c}}
\left[\Omega_0 (1+z)^3 +(1-\Omega_0)
\right]
\label{eq:crit}
\end{equation}
%
%%\begin{equation}
%%\rho_{\rm b} (z)=\rho_{0 {\rm b}} (1+z)^2 \left[\Omega_0(1+z)+(1-\Omega_0)
%%\right]
%%\label{eq:crit}
%%\end{equation}
The subscript $0$ indicates that the parameter is to be
calculated at epoch $t_0$, $H_0$ is the Hubble constant at $t_0$ and 
$\Omega_0=\frac{8 \pi G \rho_{0 {\rm c}}}{3 H^2_0}$.
%%
%%and
%%$\Omega_{\rm R}=\frac{1}{(H_0 R)^2}$, being $R$ the radius of curvature and
%%$H_0$ the Hubble constant at $t_0$.
%%
%%and
%%$\delta_{\rm v}$
%%is provided by the spherical top-hat collapse model.
The virial overdensity,
$\delta_{\rm v}$,
is provided by the spherical top-hat collapse model,
which, for the family of flat cosmologies ($\Omega_0+\Omega_{\Lambda}=1$),
gives:
%%$\delta_{\rm v}$ is:
\begin{equation}
\delta_{\rm v}(z) \simeq (18 \pi^2+82 y-39 y^2)
\label{eq:over}
\end{equation}
%
%%\begin{equation}
%%\delta_{\rm v}(z) \simeq (18 \pi^2+82 y-39 y^2)/\Omega(z)
%%\label{eq:over}
%%\end{equation}
(Bryan \& Norman 1998), where $y \equiv \Omega(z)-1$ and $\Omega(z)$ is:
\begin{equation}
%%\Omega(z)=\frac{\Omega_0(1+z)^3}{\Omega_0(1+z)^3+
%%\Omega_{\rm R} (1+z)^2+(1-\Omega_0)}
\Omega(z)=\frac{\Omega_0(1+z)^3}{\Omega_0(1+z)^3+
(1-\Omega_0)}
\label{eq:omeg}
\end{equation}
%
%%\begin{equation}
%%\Omega(z)=\frac{\Omega_0(1+z)}{(1-\Omega_0)+\Omega_0(1+z)}
%%\label{eq:omeg}
%%\end{equation}
In the limit $z \rightarrow \infty$,
%%$\Omega=1$
equation (\ref{eq:mass})-(\ref{eq:omeg})
give the corresponding quantities for 
%%reduce to those of
the SCDM model.
%%A useful formula to calculate $r_{\rm 200}$ is given by NFW97:
%%\begin{equation}
%%r_{200}=1.63 \times 10^{-2} \left( \frac{M}{h^{-1} M_{\odot}} \right)^{1/3}
%%\left( \frac{\Omega_0}{\Omega(z)}\right)^{-1/3} \frac{1}{1+z} h^{-1} kpc
%%\end{equation}

\section{Results}

In NFW96 and NFW97, the N-body simulations were interpreted by means of
a model, a step-by-step calculation of the density profile that is also useful
to calculate the mass and redshift dependence of the concentration
parameter, $c$.
This model assigns to each halo of mass $M_{\rm v}$
identified at
%epoch ...,
$z=z_0$ a collapse redshift, $z_{\rm c}$, defined as the time at which
half mass of the halo was first contained in progenitors more massive than
some fraction $f$ of the final mass, $M_{\rm v}$.
Lacey \& Cole (1993) showed that a randomly chosen mass element from a
halo having mass $M_{\rm v}$, identified at redshift $z_0$, was part of a progenitor
with mass exceeding $fM_{\rm v}$ at the earlier redshift $z$ with probability 
\begin{equation}
P(f M_{\rm v},z|M_{\rm v},z_0)=erfc \left(
\frac{\delta_{\rm crit}(z)-\delta_{\rm crit} (z_0)}
{\sqrt{2 \left(\sigma^2_0(f M_{\rm v})-\sigma^2_0(M_{\rm v}\right)}}\right)
\label{eq:erfc}
\end{equation}
where $\sigma^2_0$ is the linear variance of the power spectrum and
$\delta_{\rm crit}(z)$ is the density threshold for spherical collapse
at redshift $z$. 
The collapse redshift $z_{\rm c}$ is determined setting $P=1/2$.
%%\vskip2cm
%%Then $z_{\rm c}$
%%can be calculated using Press-Schecter formalism (Lacey \& Cole 1993):
%%\begin{equation}
%%erfc \left(
%%\frac{\delta_{\rm crit}(z_{\rm c})-\delta^0_{\rm crit}}
%%{\sqrt{2 \left(\sigma^2_0(f M_{\rm v})-\sigma^2_0(M_{\rm v}\right)}}\right)=1/2
%%\label{eq:erfc}
%%\end{equation}
%%where $\sigma^2_0$ is the linear variance of the power spectrum and
%%$\delta_{\rm crit}(z)$ is the density threshold for spherical collapse
%%at redshift $z$.
%%\vskip2cm
Assuming that the characteristic density of a halo is
proportional to the density of the universe at the corresponding $z_{\rm c}$
then we have (NFW97):
\begin{equation}
\delta_{\rm n}(z_0) = C \rho_{\rm b}(z_0) \left(
\frac{1+z_{\rm c}}{1+z_0}
\right)^3
\label{eq:delt}
\end{equation}
\begin{figure}
%\vspace{302pt}
\psfig{file=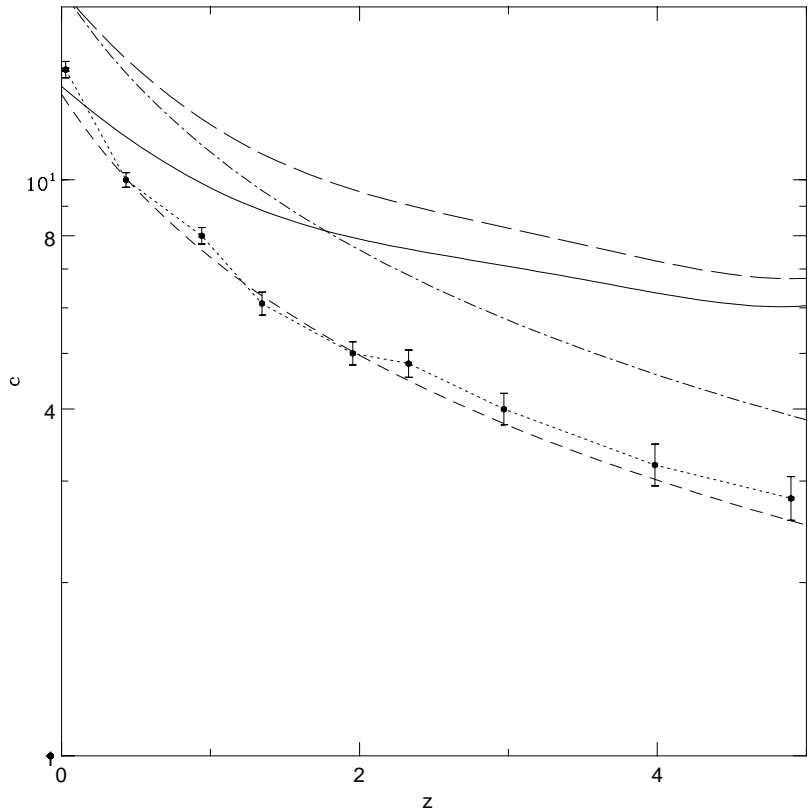,width=12cm}
\caption[]{Evolution of the concentration parameter, $c$, 
in the $\Lambda$CDM and SCDM models of the text for a halo of
$8 \times 10^{11} h^{-1} M_{\odot}$.
The solid line is the behavior of $c$ for haloes of
$8 \times 10^{11} M_{\odot}$
predicted by NFW97 analytic procedure in the case of the $\Lambda$CDM model
of the text, 
while the dotted line with errorbars
represents the evolution of $c$ obtained in B99
simulations. The short-dashed line represents the prediction for $c(z)$ 
by the present model.
In the upper two lines (long-dashed line and dot-dashed line),
I show the result of the same calculation for the SCDM model.
Similarly to the $\Lambda$CDM, NFW97 model (long-dashed line)
over-predicts the concentration, $c$, with respect to
the present model (dot-dashed line) prediction.
%%and in agreement with Bullock et al. (1999) analytic result
%%and simulations.
}
\label{Fig. 1}
\end{figure}
Given $M_{\rm v}$ and $z_0$ it is possible to obtain $z_{\rm c}$ from
equation (\ref{eq:erfc}) and $\delta_{\rm n}$ from  equation (\ref{eq:delt}), thus
completely specifying the density profile.\\
NFW97 model is in agreement with N-body simulations at $z=0$, for several
different cosmological models (NFW96; NFW97; DP2000;
B99) but
as shown by B99, it does not reproduce properly the
redshift dependence of the halo profiles as seen in their simulation:
it over-predicts the concentration, $c$, at early times, $z \ge 1$. \\
In DP2000, I showed that the improved SIM model,
introduced in that paper, gives good results in predicting the shape of
the dark haloes profiles and the mass dependence of the concentration
parameter, $c$, both in a SCDM model and in a scale-free universe.
In that paper, I did not study the redshift dependence of the concentration parameter.
Here, in order to answer this question,
I calculated the evolution of $c$ for two different cosmologies, namely
SCDM ($\Omega=1$,$h=0.5$,$n=1$; $\sigma_8=0.63$)
and $\Lambda$CDM ($\Omega_{\rm m}=0.3$, $\Omega_{\Lambda}=0.7$,
$h=0.7$, $\sigma_8=1$). The results are plotted in Fig. 1-3. \\
The solid
line of Fig. 1 represents the expected behavior for 
$8 \times 10^{11} h^{-1} M_{\odot}$ haloes as predicted from the NFW97 model
in a $\Lambda$CDM model. The dotted line with errorbars
represents the $c(z)$ median as obtained in N-body simulations by
B99 and the Poisson errorbars were obtained by the
quoted authors
from the profile fitting procedure: after identifying a centre for the
halo, they count particles in logarithmically spaced radial bins and
assign corresponding Poisson errors based on the count in each bin.
The short-dashed line is $c(z)$, obtained by means of the present model.
The model was
normalized to match the normalization of the $z=0$ relation. As can be
seen from Fig. 1, the NFW97 model (solid line)
overpredicts the concentration, $c$, by $\simeq 50\%$ at $z=1$,
with respect to the present model (short-dashed line) and by $\simeq 40\%$
with respect to that of
B99 (dotted line with errorbars).
The disagreement increases with
increasing $z$. The present model predicts
$c \propto 1/(1+z)$ in agreement with that of B99.
%
%%$c \propto 1/(1+z)$.
%%The discrepancies between the present model
%%and Bullock et al. (1999)
%%model diminishes with increasing $z$ and at $z \ge 1$ the two models
%%are in good agreement.
In the upper two lines of Fig. 1 (long-dashed line, dot-dashed line),
I show the result of the same calculation for the SCDM model.
Similarly to the case of the $\Lambda$CDM model, NFW97 model (long-dashed line)
overpredicts the concentration $c$ with respect to the model
of this paper (dot-dashed line)
%%. In the present model (dot-dashed line)
in which $c \propto 1/(1+z)$, in agreement with the analytic result and the
simulations of
B99.\\
The scaling behavior of $c$ can be explained in the same way described by
B99: in the present model the scale radius, $a$, is roughly
constant, then the $z$ dependence of $c=r_{\rm v}/a$ comes from the virial
radius $r_{\rm v}$: aside from the $z$ dependence of
$\delta_{\rm v}$, 
both in the SCDM and $\Lambda$CDM model $r_{\rm v} \propto 1/(1+z)$. \\
%%and
%%in the $\Lambda$CDM $r_{\rm v} \propto \Omega^{1/3}/(1+z)$. \\
%
%%%The rapid decline of $c$ with $z$ is due to the fact that haloes at high
%%%redshift are beginning to form and are hence more diffused.
%
Till now, I showed that the model proposed in this paper, in agreement
with claims by B99, predicts a different
redshift dependence for the concentration $c(z)$ with respect to NFW97.
It is time to discuss the reasons behind the quoted discrepancy and,
at the same time, the agreement of the results of the present paper
with the B99 paper. \\
The different result of these two models
with respect to NFW97 is fundamentally
due to the different way of defining the collapse redshift, $z_{\rm c}$:
as previously reported in this section, NFW97 defines
$z_{\rm c}(M_{\rm v},f,z_0)$ as the
time at which half the mass of the halo was first contained in progenitors
more massive than some fraction $f$ of the final mass, $M_{\rm v}$. Then the
collapse redshift is obtained through the Press-Schechter formalism,
equation (\ref{eq:erfc}) and
finally the characteristic overdensity of the halo, $\delta_{\rm n}$,
(which according to equation (\ref{eq:dn}) is connected to the concentration, c)
is assumed to be proportional to the density of the universe at $z_{\rm c}$.
As a consequence of this way of defining $z_{\rm c}$,
the NFW97 prediction for $c$ eventually goes to a constant value at high
redshift, because $z_{\rm c}$ becomes closer and closer to $z_0$, as can
be easily found from equation (\ref{eq:erfc}). In other words, at
sufficiently high redshift the halos
collapse redshift, $z_{\rm c}$, becomes essentially indistinguishable from
the redshift, $z_0$, at which they are analyzed and their concentrations tend
to a constant. \\
Things go differently with respect to the previous discussion, both 
%%, (but for different reasons as I am going to
%%explain in what follows)
in the present paper and in that of B99.
In the present paper, I use the SIM, which,
as it is well known,
%%the SIM
allows one to establish, for a given
power spectrum, a relationship between the mass $M$ and its formation
(collapse) epoch, $z_{\rm c}$ (see Peebles 1980, Section 19; Gunn \& Gott 1972;
Avila-Reese, Firmani \& Hernandez 1998). In fact, given a density perturbation,
a shell with initial comoving radius $r_{\rm i}$ and mass
$M= \frac{4 \pi}{3} \rho_{\rm b} r_{\rm i}^3 (1+ \overline{\delta_{\rm i}})$,
will expand to a maximum radius $r_{\rm m}$, given
by equation (\ref{eq:rv}) or by $r_{\rm m}= r_{\rm i}/\overline{\delta_{\rm i}}$
in an Einstein-de Sitter universe,
%%and then oscillate in and out.
and then collapse will occur. 
The time of maximum expansion in an Einstein-de Sitter universe is given by:
\begin{equation}
%%t_{\rm m}=t_{\rm c}/2=\frac{3 \pi}{4} \frac{t_0}{\overline{\delta}^{3/2}} 
t_{\rm m}=\frac{t_{\rm c}}{2}=\frac{\pi}{2 H_{\rm i}}
\frac{1}{
\overline{\delta}_{\rm i}^{3/2}}
\label{eq:tm}
\end{equation}
where $t_{\rm c}$ is the collapse time and 
$\overline{\delta_{\rm i}}$ is the mean fractional density
excess measured at time $t_{\rm i}$.
%%, assuming linear growth.
%%%
%%%In other words, from equation (\ref{eq:rv})
%%%we see that the collapse redshift $z_{\rm c}$ is function only of the 
%%%virial radius and then of the virial mass,
%%is function only
%%of the collapse redshift $z_{\rm c}$,
%%%in agreement with the assumption made
%%%by B99 (see their Eq. 8). This means that $z_{\rm c}$
%%%is uniquely determined by $M_{\rm v}$ and is independent of $z_0$ with
%%%the consequence that the behaviour of $c(z)$ to tend to a constant
%%%at high redshift, when
%%%$z_{\rm c} \rightarrow z_0$, is no more present.
%%%The implication is that $c(z) \propto 1/(1+z)$ and the slope of the relation
%%%remains constant, and it never reaches a constant.
%%%
In other words, equation (\ref{eq:rv}) and equation (\ref{eq:tm})
tell us that the collapse redshift $z_{\rm c}$
is a function only of the virial radius (and then of the virial mass,
$M_{\rm v}$), and is independent of $z_0$, in agreement with the assumption
made by B99 in their equation (8).
%%%As a consequence, the behaviour of $c(z)$ to
%%%tend to a constant at high redshift, seen in the NFW97 model, and due to the
%%%fact that at high redshift $z_{\rm c} \simeq z_0$, is no more present in
%%%the present? paper.
As a consequence, the condition $z_{\rm c} \simeq z_0$, encountered in the
NFW97 model at high redshift, implying that $c(z)$ tends to a constant value,
%%at high redshift,
is no more present in the model of this paper.
This implies that the slope of the relation $c$-$z$ remains constant.
%%and $c$ never reaches a constant??
In conclusion, $c(z) \propto 1/(1+z)$ and $c$ never tends to a constant.
The same result is found
in B99 model.
%%%
%%%I want to remark
%%%that B99 result was obtained
%%%by means of the 'ad-hoc' assumption 
%%%that the collapse redshift, $z_{\rm c}$, is uniquely determined by the
%%%virial mass, $M_{\rm v}$, and not by $z_0$ (see their equation (8)).
%%%This assumption showed
%%%'a-posteriori' that the model so modified was in good agreement with
%%%N-body simulations for what concerns the evolution of the concentrarion,
%%%$c$, with redshift $z$.
\begin{figure}
%\vspace{302pt}
\psfig{file=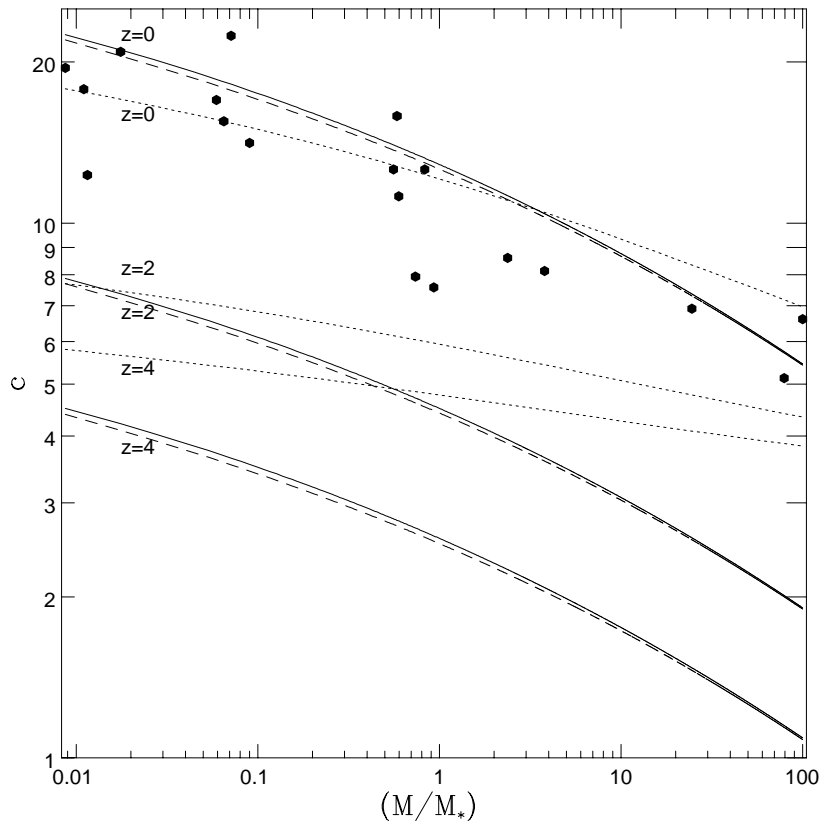,width=12cm}
\caption[]{Evolution of the concentration parameter, $c$, 
in the SCDM model of the text as function of the mass and for
different redshifts.
The solid lines represent the prediction of the model of this paper at
$z=0$, $z=2$, $z=4$. The dotted lines represent
the prediction of NFW97 for the same redshifts
and for the values of the parameters, constant in time, $f=0.01$ and
$C=3 \times 10^4$, defined in NFW96 and NFW97, while the dashed lines
represent the
predictions of B99 model for the values of the two parameters,
constant as a function of both $z$ and mass,
$F=0.01$, $K=3.8$, defined in their paper. The filled hexagons represent
the concentration $c$ at $z=0$ obtained by NFW96 and NFW97 N-body simulations.
}
\label{Fig. 2}
\end{figure}
I also would like to remark that in the present model and in that of
B99, it is possible, at high redshift,
that $z_{\rm c} < z_0$
if a halo is much more massive than the characteristic mass,
$M_{\ast}$~\footnote{I recall that as shown in DP2000 for the adopted normalization
of the CDM spectrum $M_{\ast}= 3 \times 10^{13} M_{\odot}$}, at that epoch,
that is, the characteristic collapse time for the
subclumps may take place in the future. This strange result
means that the collapse of the
sublumps happened only because the large scale structure,
within which they were sitting,
collapsed, or, in other words, the collapse of the subclumps happened at
the same time as the collapse of the whole halo itself.  In this case
the halo has a very low concentration.  This implies that
haloes much larger than $M_{\ast}$ will always be of low concentration, because their
subclumps did not collapse much earlier than the halo itself.
%%%
We may also add that
the
concentration declines dramatically with $z$ because haloes at high redshift
are just beginning to form and are therefore more diffuse.\\
%%%
A possible question that may arise at this point is why the NFW97 model
works well at $z=0$ and not at higher redshift.\\
In fact at $z=0$, the NFW97 model correctly predicts 
the mass-density
relation obtained from N-body simulations and the result is also in agreement
with B99 and DP2000.
The answer to the previous question is that the problem of NFW97,
seen at $z>0$,
is no more present at $z=0$ because at this redshift, 
the extended Press-Schechter formula, that NFW97  used to determine
$z_{\rm c}$, never gives $z_0 \simeq  z_{\rm c}$, so the problem of the wrong
prediction of $c(z)$, seen at high redshift, 
%%of the NFW97 observed at $z>0$
is no more present at $z=0$.
%%and the model gives
%%the same prediction for the mass-density relation of B99
%%and DP2000.
%%never arises.
%%%
%%%When
%%%$z_{\rm c} >> z_0$ then all the three models give very similar results.\\
%%%
%%As a final result, NFW97 obtains z_c(M,f,)
In order to further clarify the previous discussion between the differences in the
model of this paper and NFW97, I calculated the concentration $c$
as a function
of mass, $M$, in a CDM model for different values of
redshift in the case of
NFW97 model, in that of
B99 and in the model of this paper. The result is shown in
Fig. 2. The solid lines represent the prediction of the model of this paper at
three different redshifts, $z=0$, $z=2$, $z=4$. The dotted lines represent
the prediction of NFW97 for the quoted redshifts, while the dashed lines are the
predictions of B99 model. The filled hexagons represent
the concentration $c$ at $z=0$ obtained by NFW96 and NFW97 in their
N-body simulations.
%%%
%%%Fig. 2 shows that $c$ decreases with increasing mass in agreement with
%%%the idea? that small mass haloes collapse earlier than larger mass ones.
Fig. 2 shows that for a fixed mass and at $z=0$,
the evolution of $c$ as a function of mass $M$ is, as known, succesfully
predicted by NFW97 model for the values of the parameters, constant in time,
$f=0.01$ and
$C=3 \times 10^4$, defined in NFW96 and NFW97.
B99 model (for the two parameters, constant as function of $z$ and
mass, $F=0.01$, $K=3.8$, see
their paper for a definition) and the model
of this paper reproduce the $z=0$ results of NFW97 quite well over the range
$\frac{M}{M_{\ast}} \simeq 0.01-100$ (being as previously reported
$M_{\ast}= 3 \times 10^{13} M_{\odot}$).
A direct comparison of the three models however shows that the NFW97 model
prediction is shallower than the other two. The same situation was present in
the $\Lambda$CDM model, as shown in Fig. 2 of B99. Also
in this case, NFW97 predicted slope for $c$ is shallower than
B99 model which reproduce better the simulations data.
At $z=2$ the discrepancy between the NFW97 and the other two models is evident.
Fig. 2 evidently shows that, going from $z=0$ to $z=2$,
the slope of the relation $c-M$, predicted
by B99 and the model of this paper,
remains roughly constant
%%changing $z$
and moreover gives lower values of $c$ with respect to the NFW97 model.
The figure also shows that the NFW97 model for $z=2$ have a smaller slope
with respect to the $z=0$ case.
The situation now described is even more evident in the $z=4$ case.
This situation was expected because,
as previously quoted in the precedent discussion, with increasing $z$,
$z_{\rm c} \rightarrow z_0$, and one expects that $c$ goes to a constant,
in fact
we see that the slope of the $c$ relation predicted by the NFW97 models 
decreases going from $z=0$ to $z=4$.
%%In the case of the other two models, the slope of the relation
%%remains roughly constant.
In other terms, the B99 model and the model of this paper
predict that $c(z) \propto 1/(1+z)$ and that the
relation never tends to a constant for a particular high redshift, which
is the case for the NFW97 model. As previously quoted,
the reason why NFW97 model gives a succesful description of the $c-M$ relation
in the case $z=0$ is due to the fact that in this case 
never happens that $z_{\rm c} \simeq z_0$. \\
Finally, I want to remark that although B99 model gives predictions
for the $c(M,z)$ relation in good agreement with N-body simulations
and with the model of this paper, there is a fundamental difference 
between these two models. B99 model makes the
%%was build with the
'ad-hoc' assumption that $z_{\rm c}$ depends only on mass and not on
$z_0$, in order to explain the results of the N-body simulations performed
in the same paper.
The model of the present paper is a direct consequence of SIM, without additive
assumptions. \\
%%It is good that you can explain what was
%%an ad hoc assumption in Bullock et al.
%%Differenza mio modello e Bullock......
\begin{figure}
%\vspace{302pt}
%%\psfig{file=fig2.ps,width=12cm}
\psfig{file=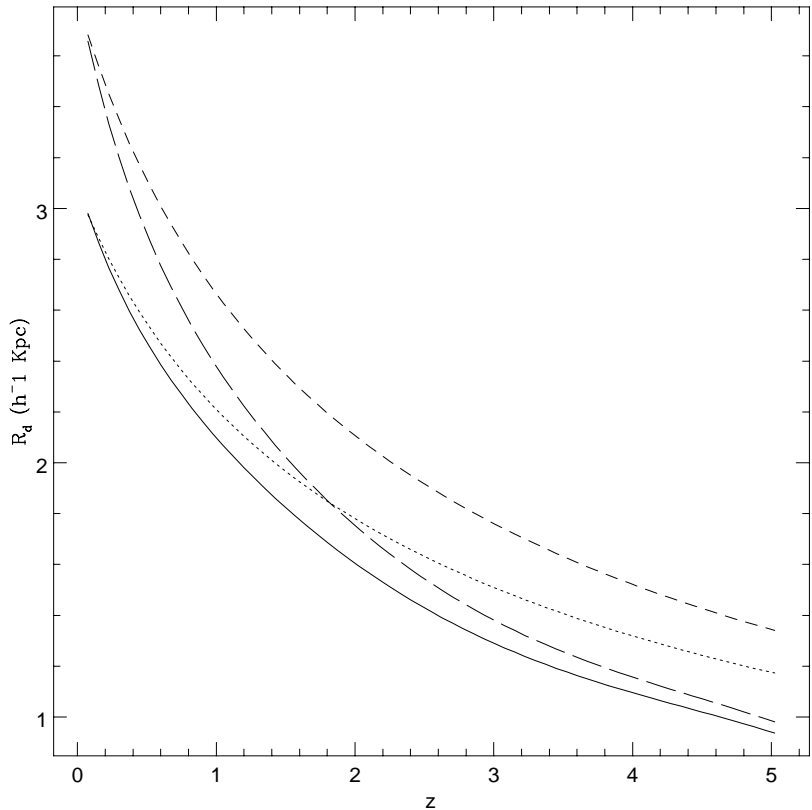,width=12cm}
\caption[]{Disk size, $R_{\rm d}$, as a function of $z$ for the SCDM and
$\Lambda$CDM model
introduced in the text 
assuming that $M_{\rm v}=8 \times 10^{11} M_{\odot}$.
The solid
and dotted line
represents $R_{\rm d}$ for the SCDM model, obtained from the B99 fitting
formula (their Eq. (7)), respectively assuming that
the evolution of $c$ follows NFW97 model (solid line)
and the model of this paper (dotted line).
%%The dotted and long-dashed line represents $R_{\rm d}$ for the
%%$\Lambda$CDM model
%%of the model presented in this paper for the SCDM model.
The long-dashed and short-dashed line
represents $R_{\rm d}$ for a $\Lambda$CDM model
obtained from the B99 fitting formula respectively assuming that
the evolution of $c$ follows NFW97 model (long-dashed line)
and the model of this paper (short-dashed line).
%%. The short-dashed
%%line is the prediction of the model presented in this paper.
}
\label{Fig. 3}
\end{figure}
Now I want to discuss some consequences of the results
obtained for c(z) on galaxy modelling at high redshift. The discussion
is intended to gain only a qualitative understanding of how $c(z)$ may
affect some features of galactic formation (a more complete discusion
is given in B99).\\
As pointed out by B99, one expects that 
the described behavior of $c(z)$ should have a large impact on galaxy
modelling at high redshift and for interpreting high redshift data
(e.g., evolution of the Tully-Fisher relation (Vogt et al. (1997) and
the nature of Lyman Break Galaxies (Steidel et al. 1996)).
%%%
%%%\begin{figure}
%%%\psfig{file=Icdm.ps,width=12cm}
%%%\caption[]{The ratio of the disk brightness $I$(SIM), obtained assuming
%%%that $c(z)$ is given by the model of this paper, and $I$(NFW), obtained
%%%assuming that $c(z)$ follows NFW97 model. The calculation is done for the
%%%SCDM model introduced in the text.
%%%}
%%%\label{Fig. 5}
%%%\end{figure}
%%%
According to the standard picture of galaxy formation, structures grow
hierarchically from small, initially Gaussian density fluctuations.
Collapsed, virialized dark matter haloes condense out of the initial
fluctuation field. Gas associated with such dark haloes cools and condenses
within them, eventually forming galaxies.
In this scenario, the growth of the dark matter haloes is not much affected
by the baryonic components, but determines how they are assembled into
nonlinear units. The halo density profile determines many of the properties
of galaxy disks, e.g. their size and surface brightness.
%
%%%According to the standard picture of galaxy formation,
%%%dark matter haloes form hierarchically bottom-up via
%%%gravitational amplification of initial density fluctuations.
%%%The haloes carry with them
%%%gas whose eventual cooling and contraction leads to the formation of
%%%luminous disk galaxies at the centres of the haloes. The halo profile affects
%%%both gas cooling and infall and several fundamental properties of the disk, such as
%%%size and surface brightness.
%
In order to show the effect of $c$ evolution
on disk properties, it is possible to use
a fitting formula given by Mo, Mao \& White (1998) based on several
assumptions about the halo and disc make-up and depending on some free
parameters, or
more easily the fitting
formula introduced in B99 (Eq. (7)):
\begin{equation}
R_{\rm d}=5.7 h^{-1} {\rm Kpc}
\left(
\frac{r_{\rm v}}{100 h^{-1} {\rm Kpc}}
\right)
\left[1+\left(c(z)/3.73\right)^{0.71}
\right]^{-1}
\end{equation}
which for $1<c<50$ is good within $1\%$.
%%as done by B99.
In Fig. 3,
%%and Fig. 4
%%%
%%%\begin{figure}
%%%\psfig{file=Ilcdm.ps,width=12cm}
%%%\caption[]{Same as Fig. 5 but for the $\Lambda$CDM model
%%%of the text}
%%%\label{Fig. 6}
%%%\end{figure}
%%%
I plot the $z$ dependence of the disk size, $R_{\rm d}$,
%%respectively
in the SCDM and $\Lambda$CDM model,
assuming that $M_{\rm v}=8 \times 10^{11} h^{-1} M_{\odot}$,
as assumed also in Fig. 1.
In both cases the lower concentration of haloes at high redshift produces
a disk size larger
%%(dashed line)
than that obtained using the NFW97 model for the evolution of $c$.
%%%%Mo et al. (1997) model.
In fact in the $\Lambda$CDM model taking account of the $c-z$ dependence
found in this paper, I find the short-dashed line, while the long-dashed
line is obtained with the NFW97 $c-z$ dependence. A similar situation
is found in the SCDM model: using the $c-z$ dependence of this paper, I
obtained the 
dotted line while using that of NFW97 I obtained the solid line.
%%%
%%%The dependence of the surface brightness, $I(z)$, from $c$ and $z$ can be
%%%obtained remembering that it scales as $I \propto R_{\rm d}^{-2}$. As shown in
%%%Fig. 5 and 6, both in the SCDM and $\Lambda$CDM model the ratio of the brightness
%%%$I$(SIM), obtained using the $c(z)$ given by the present model with 
%%%$I$(NFW), obtained using the $c(z)$ given by NFW model,
%%%is smaller than 1 and decreases with $z$.
I want to stress that the last result must be taken with caution, since
it strongly dependes on the precision of the fitting formula used and on
its free parameters (if it contains any).
In fact comparing Fig. 1 and Fig. 3, it is immediately evident that,
while the concentration found in the model of the present paper
differs from that predicted by NFW97 evolution quite a lot, the differences
is reduced when we deals with $R_{\rm d}$.
%
%for example, at $z=4$, the difference obtained between the value of
%$R_{\rm d}$ calculated
%by using the NFW97 model for the evolution of $c$ and that of the
%present paper, is smaller with respect to that observed in the case of the
%concentration, $c$ (see Fig. 1).
If I had used the fitting formula of Mo, Mao \& White (1998),
to predict the variation of $R_{\rm d}$ with redshift, I should have found
that small variations in the parameters of that model, at a
level of 10\%, would have produced change in $R_{\rm d}$ larger than that
due to the redshift evolution of c.
So the validity of the last result, and that regarding
the surface brightness, $I(z)$, described in the following, is entirely based upon
the reliability of the B99 fitting formula.

Since the surface brightness, $I(z)$, scales as:
\begin{equation}
I(z) \propto R_{\rm d}^{-2}(z)
\label{eq:bri}
\end{equation}
both in the SCDM and $\Lambda$CDM model, we may expect that
this parameter
is also modified by the strong evolution of $c$ with redshift. Using Fig. 3 and
equation (\ref{eq:bri}) it is easy to find that 
%%%
%%%the ratio of the brightness
%%%$I$(SIM), obtained using the $c(z)$ given by the present model with 
%%%$I$(NFW), obtained using the $c(z)$ given by NFW model,
%%%is smaller than 1 and decreases with $z$.
%%%
%%%This means that
according to the model of the present paper 
disks are dimmer with respect to NFW prediction and
that the effect
increases with increasing $z$. 
Besides the two examples that I have just given,
the strong evolution of the concentration parameter, $c$, affects
several other properties of disks and, as previously reported, has some
implications for galaxy modelling. For example, the shape of a disk
rotation curve depends, among other, on the concentration, $c$,
of its halo.
Since more strongly peaked curves are found in more concentrated haloes
%Since
and evolution produces a reduction of the concentration of haloes
one expects that evolution tends to produce less peaked rotation curves.
Further implications of the $c(z)$ dependence shown in this paper are
discussed in B99. However, as stressed by the quoted
author, much work remains to be done in order to have a deeper understanding
of  the implications
of the results obtained in this and their paper on galaxy modelling. 
%
%%%Although I showed some implications of the model
%%%for the evolution of dark matter haloes profiles,
%%%there is much work to be done in order to have a deeper
%%%understanding of  the implications
%%%of the result obtained in this paper on galaxy modelling. The subject
%%%is so important and interesting that it deserves to be studied carefully
%%%in some future papers. 
%
%%much work remains to do
%%in future papers.
%%In particular the maximum of the circular velocity increases with increasing
%%$c$ (Bullock et al. 1999) and consequently 
%%As a consequence,
%%lower concentration halos have smaller values of $V_{\rm max}$ and this
%%leads to the formation of larger and dimmer galaxies (Bullock et al. 1999).
%%Moreover low concentration halos will hinder cooling. 

\section{Conclusions}

In this paper I studied the evolution of the concentration parameter, $c(z)$,
for fixed mass haloes,
in the SCDM and $\Lambda$CDM model
by means of the improved SIM introduced in
DP2000.
%
%%The evolution of $c(z)$ depends on the
%%cosmological model chosen.
The results of the paper can be summarised as follows:\\
a) Both in the SCDM and $\Lambda$CDM model
the evolution of $c(z)$ is much stronger than that expected
from the NFW97 model: in the model of the present paper
$c(z) \propto 1/(1+z)$, while NFW97 model overpredicts
the concentration $c$ by $50\%$ at $z=1$, with respect to the model
of the present
paper, and by $40\%$, with respect to that of B99. NWF97 predicts that the
concentration $c$ tends to a constant value for large $z$, so
the quoted disagreement increases with increasing $z$. \\
%%while in a $\Lambda$CDM model $c(z) \propto \simeq \Omega^{1/3}/(1+z)$.
%%In both cases,
b) A comparison of the results of the present paper
with the high-resolution N-body simulations
of B99
%(they performed simulations
%only for the $\Lambda$CDM model)
shows a good agreement both for the SCDM and $\Lambda$CDM model. \\
%%In the case of the SCDM model the result
%%of our model coincides with the analytic model and
%%N-body simulations of Bullock et al. (1999). \\
c) The different redshift dependence of $c(z)$, obtained in the present paper
and B99, with respect to NFW97, is fundamentally
due to the different way of defining the collapse redshift, $z_{\rm c}$:
as a consequence of the way of defining $z_{\rm c}$ in NFW97, at
sufficiently high redshift the halos
collapse redshift, $z_{\rm c}$, becomes essentially indistinguishable from
the redshift, $z_0$, at which the halos itself are analyzed and their
concentrations tend to a constant. \\
d) The present model has an advantage with respect to B99, namely:
B99 results are obtained assuming the 'ad-hoc' assumption that
$z_{\rm c}$ depends only on mass, and not on $z_{\rm o}$, while the results of
the present paper are direct consequence of SIM, without additive assumptions.\\
e) The result has important consequences on galaxy modelling at high
redshift and on some disks characteristics.
In particular the disk size obtained in the present model is larger than that
obtained using the B99 fitting formula together with NFW prediction for $c(z)$. The reverse
is true for the disk brightness.     \\
%%The investigation of these topics and the study of other connected to the
%%subject
%These are important topics
%constitutes material of study for future papers.  \\
\section*{Acknowledgements}
I would like to thank Prof. E. Recami, Dr. Y. Eksi and Dr. J. Bullock
for some useful comments that helped me in improving this paper.
Special thanks go to E. Nihal Ercan for her continous support, 
encouragement and for her special and rare humanity.   
%A. Del Popolo and E.Nihal Ercan would like to thank
%Bo$\breve{g}azi$\c{c}i University
%Research Foundation for the financial support through the project code
%01B304.
{}

\bsp

\label{lastpage}

\end{document}